 \documentclass[final,3p,times,twocolumn]{elsarticle}

\usepackage{graphicx}

\usepackage{amssymb}

\usepackage{dcolumn}
\usepackage{amsmath}
\usepackage{bm}

\journal{Solid State Communications}

\begin{document}

\begin{frontmatter}



\title{Spin interactions, relaxation and decoherence in quantum dots}

\author{Jan Fischer}
\address{Department of Physics, University of Basel, Klingelbergstrasse 82, CH-4056 Basel, 
  Switzerland}

\author{Mircea Trif}
\address{Department of Physics, University of Basel, Klingelbergstrasse 82, CH-4056 Basel, 
  Switzerland}

\author{W.~A. Coish}
\address{Institute for Quantum Computing and Department of Physics and Astronomy,
University of Waterloo, 200 University Ave. W., Waterloo, ON, N2L 3G1,
Canada}

\author{Daniel Loss}
\address{Department of Physics, University of Basel, Klingelbergstrasse 82, CH-4056 Basel, 
  Switzerland}

\begin{abstract}
We review recent studies on spin decoherence of electrons and holes in quasi-two-dimensional 
quantum dots, as well as electron-spin relaxation in nanowire quantum dots. The spins of confined 
electrons and holes are considered major candidates for the realization of quantum 
information storage and processing devices, provided that sufficently long 
coherence and relaxation times can be achieved. The results presented here indicate
that this prerequisite might be realized in both electron and hole quantum dots,
taking one large step towards quantum computation with spin qubits.
\end{abstract}


\end{frontmatter}



\section{Introduction}\label{sec:intro}
%
%

Quantum bits are the quantum-mechanical counterparts of classical bits -- the
basic building blocks for quantum information processing devices and, eventually,
for a quantum computer. 
In general, a quantum bit (`qubit') can be regarded as a quantum-mechanical two-level
system, its states representing the logical computational states $|0\rangle$ and $|1\rangle$.
In contrast to their classical counterparts, quantum computers (or rather quantum
algorithms) make direct use of quantum phenomena, such as superpositions and
entanglement, which has been shown theoretically to significantly speed up many 
computational tasks, such as prime factorization \cite{Shor1997} or searching 
\cite{Grover1997}. The physical realization of quantum-information storage and processing
devices, as well as the implementation of gates to perform computational operations 
on qubits, has yet to be shown on a large enough scale to actually perform viable
quantum computation. There is still debate on the `best choice' of a quantum-mechanical
two-level system to use as a qubit.

One promising candidate for the realization of a single qubit is an electron which
is confined to a semiconductor quantum dot
and whose spin states ${|\uparrow\rangle}$ 
and ${|\downarrow\rangle}$ represent the logical qubit states required for quantum 
computation \cite{Loss1998}. 
Such a qubit is also referred to as a `spin qubit'.

One significant obstacle towards the realization of quantum information
storage and processing devices based on spin qubits is the limited lifetime 
of information encoded in the electron 
spin states. The electron is not isolated from its environment but is coupled
to the nuclei of the semiconductor host material, via both lattice-mediated 
spin-orbit interactions and interactions with the nuclear spins.
The coupling between the electron and the nuclei influences the electron spin state
and, hence, changes the quantum information that has been initialized on the spin qubit.
On the other hand, it provides a way to control the electron-spin state 
(see Sec. \ref{sec:spinorbit}).
It is therefore of primary interest to study and understand the nature of spin 
interactions in quantum dots.

There are two processes that lead to a limited lifetime of information stored in
spin qubits. On the one hand there is \emph{relaxation}, 
i.~e., the transition from the excited state $|\uparrow\rangle$ into the ground state 
$|\downarrow\rangle$, which happens on a characteristic timescale $T_1$. \
On the other hand, spin-state superpositions decay on a timescale $T_2$, and the 
accordant process is referred to as \emph{decoherence}. Although
relaxation inevitably also leads to decoherence, these two processes are not equivalent.
Decoherence can also occur without changing the population of the spin states, a phenomenon
which is known as \emph{pure dephasing}.

The main physical mechanisms that lead to relaxation and decoherence of electron spin states
in quantum dots are spin-orbit interactions and nuclear-spin interactions.
The first couples the electron
spin to its orbital momentum via the electric field created by the nuclei.
The second couples the electron spin (and orbital angular momentum) to a fluctuating magnetic 
field created by the nuclear spins. 

More recently, proposals have been put forward to use quantum-dot-confined heavy holes for 
quantum-information storage devices rather than
electrons, since hole spins are believed to interact less strongly with the surrounding nuclei.
It has been shown both theoretically and experimentally that hole spins interact 
very differently with their nuclear environment than electrons.
While the timescales $T_1$ for relaxation may be comparable for electrons and holes,
typical hole-spin coherence times $T_2$ seem to be significantly longer than 
for electrons.

Here we give an overview of recent research on spin
interactions and spin dynamics of quantum-dot-confined electrons and holes. 
The paper is organized as follows:
In Sec. \ref{sec:mechanisms}, we derive the interactions of an electron spin with
the nuclei in a quantum dot. 
In Sec. \ref{sec:hyperfine:electron}, we review recent results on hyperfine-induced 
decoherence of electrons in quantum dots.
In Sec. \ref{sec:hyperfine:hole}, we discuss the interaction of a confined heavy hole
with the nuclear spins in a quantum dot and the resulting hole-spin decoherence.
In Sec. \ref{sec:spinorbit}, spin-orbit interactions of an electron confined to a
nanowire quantum dot and the resulting spin relaxation are treated.
Finally, we summarize in Sec. \ref{sec:summary}.

\section{Spin interactions in quantum dots}\label{sec:mechanisms}
%
%

In this Section, we give a microscopic derivation of the spin-orbit and nuclear-spin interactions. 
The interaction of a relativistic electron with the electromagnetic field created by a nucleus 
is described by the Dirac Hamiltonian
\begin{equation}
  \label{eq:dirachamiltonian}
  H_D = \mathbf{\alpha} \cdot \mathbf{\pi} + \beta m c^2 + qV,
\end{equation}
where $m$ is the electron rest mass, $q=-|e|$ is the electron charge,
$\mathbf{\pi} = c(\mathbf{p} - q \mathbf{A})$, $c$ is the speed of light,
$\mathbf{p}$ is the momentum, $V$ and $\mathbf{A}$ are the scalar
and vector potential of the electromagnetic field induced by the nucleus, and
\begin{equation}
  \mathbf{\alpha} = \begin{pmatrix}
  0 & \mathbf{\sigma}\\ \mathbf{\sigma} & 0
  \end{pmatrix},
  \quad
  \mathbf{\beta} = \begin{pmatrix}
  \mathbf{1} & 0\\ 0 & {-\mathbf{1}}
  \end{pmatrix}
\end{equation}
are the $4\times4$ Dirac matrices with $\mathbf{\sigma}$ being the vector of Pauli matrices and
$\mathbf{1}$ the $2\times2$ identity matrix.

The Dirac Hamiltonian \eqref{eq:dirachamiltonian} acts on a 4-spinor $\psi = (\chi_1, \chi_2)^t$,
where $\chi_1$ and $\chi_2$ are 2-spinors describing the electron and the positron, respectively.
Using this notation, the Dirac equation $H_D \psi = E \psi$, with $E = mc^2 + \epsilon$, may be written 
as a pair of coupled equations for the $\chi_j$:
\begin{eqnarray}
  \label{eq:dirac-chi1}
  (\epsilon - qV) \> \chi_1 - \mathbf{\sigma} \cdot \mathbf{\pi} \> \chi_2 &=& 0,\\
  \label{eq:dirac-chi2}
  -\sigma \cdot \pi \> \chi_1 + (2mc^2 - qV + \epsilon) \> \chi_2 &=& 0.
\end{eqnarray}
Isolating $\chi_2$ in Eq. \eqref{eq:dirac-chi2} and inserting into Eq. \eqref{eq:dirac-chi1}
yields the following eigenvalue equation for the electron:
\begin{equation}
  \label{eq:chi1-relativistic}
  \left( \mathbf{\sigma} \cdot \mathbf{\pi} \> \frac{1}{2mc^2 - qV + \epsilon}
  \> \mathbf{\sigma} \cdot \mathbf{\pi} + q V \right) \, \chi_1 = \epsilon \chi_1.
\end{equation}
In the non-relativistic limit $(\epsilon - qV)/mc^2 \rightarrow 0$, $\chi_1$ and $\chi_2$
decouple and Eq. \eqref{eq:chi1-relativistic} reduces to the Pauli equation $H_P \chi_1 = \epsilon \chi_1$
with the Pauli Hamiltonian
\begin{equation}
  \label{eq:pauli}
  H_P =	\frac{1}{2m} \> (\mathbf{p} - q\mathbf{A})^2 - \frac{q \hbar}{2m} \> (\mathbf{\nabla} \times
  \mathbf{A}) \cdot \mathbf{\sigma} + qV.
\end{equation}

In general, one has to take into account the relativistic effect of a coupling between the electron
and positron 2-spinors. It is, however, possible to systematically decouple $\chi_1$ and $\chi_2$ in orders
of $1/mc^2$ by successively applying unitary transformations to the 
Dirac Hamiltonian \eqref{eq:dirachamiltonian}.
This method takes into account relativistic corrections to the Pauli
equation and is known as the \emph{Foldy-Wouthuysen transformation}.
In lowest order, this method leads to an eigenvalue equation $H_{\mathrm{FW}} \chi_1 = \epsilon \chi_1$
for the electron spinor, where $H_{\mathrm{FW}}$ contains the Pauli Hamiltonian and the
first relativistic corrections:
\begin{align}
  H_{\mathrm{FW}} = H_P - \frac{q \hbar}{4m^2c^2} \> \left( \mathbf{E} \times \frac{\mathbf{\pi}}{c}
  \right) \cdot \mathbf{\sigma} - \frac{q \hbar^2}{8m^2c^2} \> \mathbf{\nabla} \cdot \mathbf{E},
\end{align}
where we have introduced the electric field $\mathbf{E} = -\mathbf{\nabla} V$.

The terms of interest are those that couple the nuclei (giving rise to $\mathbf{E}$ 
and $\mathbf{A})$ through their charge and magnetic moment to the electron 
(with spin $\mathbf{\sigma}$ and momentum $\mathbf{p}$). These terms
are
\begin{eqnarray}
  \label{eq:spinorbit}
  H_{\mathrm{so}} &=& -\frac{q \hbar}{4m^2c^2} \> (\mathbf{E} \times \mathbf{p}) \cdot 
  \mathbf{\sigma},\\
  \label{eq:fermicontact}
  H_{\mathrm{ihf}} &=& \frac{q^2 \hbar}{4m^2c^2} \> (\mathbf{E} \times \mathbf{A}) \cdot 
  \mathbf{\sigma},\\
  \label{eq:anisotropic}
  H_{\mathrm{ahf}} &=& -\frac{q \hbar}{2m} \> (\mathbf{\nabla} \times \mathbf{A}) \cdot 
  \mathbf{\sigma},\\
  \label{eq:angular}
  H_{\mathrm{ang}} &=& -\frac{q}{m} \> \mathbf{A} \cdot \mathbf{p},
\end{eqnarray}
and are referred to as spin-orbit interaction, isotropic hyperfine interaction,
anisotropic hyperfine interaction, and the coupling of electron orbital angular momentum
to the nuclear spin, respectively.

The spin-orbit interaction manifests itself via a splitting of the electron
spin states at $\mathbf{k} \neq 0$,
even at zero external magnetic field, $\mathbf{B}_{\mathrm{ext}}=0$.
This splitting can be due to bulk inversion asymmetry (BIA) of the crystal,
or due to structure inversion asymmetry (SIA) caused, e.g., by a confinement potential
\cite{winkler}.
In III-V semiconductor quantum dots, which we will consider throughout this article,
both effects can be relevant (see Sec. \ref{sec:spinorbit}): 
the zincblende-type crystal structure of III-V semiconductors
lacks a center of inversion symmetry, causing BIA, while the strong asymmetric 
confinement of the quantum dot leads to SIA.

The importance of nuclear-spin interactions strongly depends on the system under 
consideration. For most quantum dots at low temperatures, the main mechanism leading
to spin decoherence of electrons is the isotropic hyperfine interaction \eqref{eq:fermicontact}
-- see Sec. \ref{sec:hyperfine:electron} --,
while for holes the anisotropic hyperfine interaction \eqref{eq:anisotropic} and
the coupling of electron orbital angular momentum \eqref{eq:angular} are most relevant 
-- see Sec. \ref{sec:hyperfine:hole}.
It is convenient to replace the nuclear-spin interactions \eqref{eq:fermicontact}
- \eqref{eq:angular} by equivalent effective Hamiltonians of the form 
\cite{stoneham, abragam}
\begin{eqnarray}
  \label{eq:isotropichyperfine}
  H_{\mathrm{ihf}}^{\mathrm{eff}} &=& \frac{\mu_0}{4 \pi} \> \frac{8 \pi}{3} \> 2 \mu_B 
  \gamma_N \delta(\mathbf{r}) \mathbf{S} \cdot \mathbf{I},\\
  \label{eq:anisotropichyperfine}
  H_{\mathrm{ahf}}^{\mathrm{eff}} &=& \frac{\mu_0}{4 \pi} \> 2 \mu_B \gamma_N 
  \frac{3 (\mathbf{n} \cdot \mathbf{S}) (\mathbf{n} \cdot \mathbf{I}) - \mathbf{S} \cdot 
    \mathbf{I}}{r^3},\\
  \label{eq:orbitalcoupling}
  H_{\mathrm{ang}}^{\mathrm{eff}} &=& \frac{\mu_0}{4 \pi} \> 2 \mu_B \gamma_N 
  \frac{\mathbf{L} \cdot \mathbf{I}}{r^3}.
\end{eqnarray}
Here, $\mu_0$ is the vacuum permeability, $\gamma_N = g_N \mu_N$
is the nuclear gyromagnetic ratio, $g_N$ is the nuclear g-factor, 
$\mu_B$ ($\mu_N$) is the Bohr (nuclear) magneton, $\mathbf{r}$ is the
vector pointing from the nucleus to the electron, $r=|\mathbf{r}|$, $\mathbf{n} =
\mathbf{r}/r$, $\mathbf{S}$ ($\mathbf{I}$) is the electron (nuclear) spin operator,
and $\mathbf{L}$ is the electron orbital angular momentum operator.
\footnote{Eqs. \eqref{eq:isotropichyperfine} - \eqref{eq:orbitalcoupling} describe the
coupling to a \emph{single} nuclear spin. In general, the electron couples to many nuclear
spins, resulting in a sum over all nuclei in the above equations -- see Eqs. \eqref{eq:Zeeman-hf}
and \eqref{eq:hole-ising}.}

\section{Hyperfine interaction and electron-spin decoherence}\label{sec:hyperfine:electron}

We include the Zeeman terms for the electron and nuclear spins in an applied magnetic field $B$.
For a single electron in the ground-state orbital of a quantum dot in an $s$-type conduction band, only the contact term (Eq. (\ref{eq:isotropichyperfine})) contributes appreciably to the electron-nuclear interaction, giving
\begin{equation}\label{eq:Zeeman-hf}
 H = \gamma_S BS^z + B\sum_k\gamma_k \mathbf{I}_k+\mathbf{S}\cdot\mathbf{h},
\end{equation}
where $\mathbf{h}=\sum_k A_k^e \mathbf{I}_k$ gives the nuclear field operator, where the coupling constant to the nucleus at position $\mathbf{r}_k$ is $A_k^e=v_0A_e^{j_k}|\psi(r_k)|^2$ with atomic volume $v_0$, total hyperfine coupling constant $A_e^{j_k}$ for the nucleus of species $j_k$ at site $k$, and electron envelope wavefunction $\psi(r)$.  It is useful to define the weighted average hyperfine coupling constant $A_e=\sum_j\nu_j A_e^j$, where $\nu_j$ is the abundance of isotopic species $j$.  In GaAs, $A_e\simeq 90\,\mu eV$ \cite{Paget1977}, corresponding to an effective field $I A_e/\gamma_S\simeq 5\,\mathrm{T}$ for a fully-polarized nuclear-spin system
($I=3/2$ for all isotopes in GaAs). In a quantum dot, typically $N\sim10^4-10^6$ nuclear spins have appreciable overlap with the electron wavefunction, whereas each nuclear spin only `feels' a fraction $1/N$ of the electron, leading to an asymmetry in the typical time scales for evolution of the electron and nuclear spin systems.  In the presence of a random nuclear field, the electron spin looses phase coherence on a time scale \cite{Khaetskii2002,Khaetskii2003,Merkulov2002,Coish2004} $\tau_\mathrm{el}\sim 1/\sigma_e$, where $\sigma_e^2 = \langle \mathbf{h}^2 \rangle-\langle \mathbf{h} \rangle^2$ gives the variance in the nuclear field.  The nuclear spin system looses phase coherence on a comparatively much longer time scale $\tau_{\mathrm{n}}$, due to a distribution in coupling strengths $A_k^e$ (we neglect nuclear dipolar interactions in this discussion, which may become important in large dots, where the $A_k^e$ are relatively smaller).  It is straightforward to estimate these time scales for a fully randomized nuclear spin system, giving
\begin{equation}\label{eq:timescales}
 \tau_\mathrm{el}\sim \frac{1}{\sigma_e}\sim \frac{\sqrt{N}}{A_e}, 
 \quad \tau_\mathrm{n}\sim \frac{N}{A_e}.
\end{equation}
A typical lateral GaAs quantum dot contains on the order of $N\sim 10^6$ nuclear spins, resulting in $\tau_\mathrm{el}\sim 10^{-8}\,\mathrm{s}$ and $\tau_\mathrm{n}\sim 10^{-6}\,\mathrm{s}$.


Historically, discussions of spin decoherence have often relied on Bloch-Redfield theory \cite{Bloch1957,Redfield1957}.  The results of Bloch-Redfield theory are simple; the longitudinal and transverse components of spin decay exponentially to their equilibrium values on time scales $T_1$ and $T_2$, respectively (see the introduction to this article).  However, Bloch-Redfield theory relies on two common approximations: (1) weak coupling between system and environment and (2) Markovian evolution (a short correlation time $\tau_c$ in the environment compared to the system decay time $\tau_s$).  Due to the large typical hyperfine coupling strengths in semiconductor quantum dots and the unfavorable asymmetry between bath and system decoherence times (see Eq. (\ref{eq:timescales})), a new approach is required to determine transverse-spin decoherence due to hyperfine interaction in quantum dots.

The transverse-spin decay time, determined by the inverse nuclear-field fluctuations, is relatively short, given by the time scale $\tau_\mathrm{el}$, estimated in Eq. (\ref{eq:timescales}) for a random nuclear spin system.  This mechanism leads to a rapid Gaussian decay of single spins \cite{Merkulov2002,Khaetskii2002,Khaetskii2003,Coish2004}, or two-spin states without exchange \cite{Petta2005,Churchill2008b}, and a power-law decay of two-electron states in the presence of exchange \cite{Coish2005,Klauser2006,Laird2006} and for single spins undergoing driven Rabi oscillations \cite{Koppens2007,Rashba2008}.  Fortunately, $\tau_\mathrm{el}$ is dependent on the initial conditions of the nuclear-spin system and can therefore be increased by modifying the initial conditions in the slowly-varying nuclear-spin bath to reduce the fluctuations $\sigma_e$ in the nuclear field and extend the associated decay time.  

One means of suppressing nuclear fluctuations is to polarize the nuclear-spin system.  A nuclear-spin system with significant polarization $p$ leads to a small electron spin-flip probability $P_{\uparrow\to\downarrow}\sim 1/p^2 N$ \cite{Burkard1999}.  However, the transverse-spin decay time was shown to be relatively much less sensitive to polarization: $\tau_\mathrm{el}(p) = \tau_\mathrm{el}(0)/\sqrt{1-p^2}$ \cite{Coish2004}, requiring a polarization on the order of 99\% to achieve a factor of 10 increase in $\tau_\mathrm{el}$.  The largest polarizations achieved in quantum dots by dynamic polarization through optical pumping or electron current are on the order of 40\%-60\% \cite{Bracker2005,Baugh2007,Maletinsky2009} and so in some cases, achieving large nuclear polarizations may not be practical.  Fortunately, theory suggests that a high degree of dynamical polarization should be achievable under appropriate conditions \cite{Imamoglu2003,Christ2007} and recent work indicates that it may be possible to fully polarize the nuclear spins in contact with an electron system in two dimensions \cite{Simon2007,Simon2008} or in a carbon nanotube \cite{Braunecker2008} through a phase transition due to a ferromagnetic interaction between nuclear spins mediated by the electrons.  A large polarization in the nuclear system may then be sustained for long times by performing a sequence of measurements on the nuclear system (quantum Zeno effect) \cite{Klauser2008}.

In situations where polarization is impractical, it is still possible to 'narrow' the distribution in the nuclear system (reduce the fluctuations $\sigma_e$) \emph{independent} of the nuclear polarization through, e.g., measurement \cite{Coish2004}.  There have been several proposals to achieve this narrowing in experimentally viable systems through measurement \cite{Klauser2006,Stepanenko2006,Giedke2006} or dynamical cooling \cite{Rudner2007}.  Recent experiments have shown a related narrowing in ensembles of dots under pulsed optical excitation \cite{Greilich2006,Greilich2007}, under a sequence of pulses applied to gated dots \cite{Reilly2008}, and under magnetic resonance \cite{Vink2009}.

Even when the nuclear-spin system is perfectly narrowed (corresponding to the creation of an eigenstate of the operator $h^z$, the $z$-component of the nuclear field operator $\mathbf{h}$), there will still be electron-spin decoherence due to quantum fluctuations. To estimate the effects of these quantum fluctuations is a significant technical challenge: a weak-coupling expansion for the strongly coupled electron-nuclear system is generally not possible.  This realization has led to alternative methods for finding the electron-spin dynamics, including exact solutions for a fully-polarized nuclear spin system \cite{Khaetskii2002,Khaetskii2003} and a variety of mean-field approximations \cite{Merkulov2002,Erlingsson2004,Yuzbashyan2005,Rashba2008}.  These methods are, however, restricted in their applicability \cite{Coish2007b}, so a great deal of work over the last few years has been devoted to consistent perturbative theories of electron spin decay in this system.

For a narrowed initial nuclear spin state, it is possible to derive an exact equation of motion for the transverse components of the electron spin $\langle S_+ \rangle=\langle S_x \rangle+i\langle S_y \rangle$ (assuming zero nuclear polarization) \cite{Coish2004}
\begin{equation}\label{eq:Transverse-spin-eom}
 \frac{d}{dt}\left<S_+\right>_t = i\gamma_S B\left<S_+\right>_t-i\int_0^t dt' \Sigma(t-t')\left<S_+\right>_{t'}.
\end{equation}
Here, $\Sigma(t)$ is the bath correlation function (memory kernel).  While the \emph{form} of Eq. (\ref{eq:Transverse-spin-eom}) is exact, it turns out to be a nontrivial problem to find a simple closed-form expression for $\Sigma(t)$, forcing us to resort to some approximation scheme.  Although a weak-coupling expansion is not possible for this problem, it is possible to expand the memory kernel $\Sigma(t)$ in powers of $A_e/\gamma_SB < 1$ in a large magnetic field $B$: $\Sigma(t) = \sum_k \Sigma^{(k)}(t)$.  At leading (zeroth) order in $A_e/\gamma_SB$, $\Sigma^{(0)}(t)=0$ resulting in no decay for a narrowed state, from Eq. (\ref{eq:Transverse-spin-eom}).  The first nontrivial corrections in $A_e/\gamma_SB$, $\Sigma^{(1)}\propto A_e/\gamma_SB$, give rise to an incomplete power-law decay of $\left<S_+\right>$ \cite{Coish2004}, in agreement with an exact solution for full polarization \cite{Khaetskii2002} in the limit $p=1$.  However, still further corrections in $A_e/\gamma_SB$ result in complete decay of $\left<S_+\right>$, with long-time power-law tails for polarization $|p|<1$ \cite{Deng2006,Deng2008}.  The behavior of $\left<S_+\right>$ at times long compared to the initial decay (due to $\Sigma^{(1)}$), but short compared to the long-time tails reported in Refs. \cite{Deng2006,Deng2008} has also been investigated, giving a quadratic shoulder, which can be approximated as a Gaussian \cite{Yao2006,Liu2007}.

These studies have found the decay of $\left<S_+\right>$ at long and short times to be distinctly non-exponential, suggesting non-Markovian dynamics.  However, there are good physical reasons to expect a Markovian regime for the decay of $\left<S_+\right>$.  Specifically, the system decay time is magnetic-field dependent since it depends on the bath correlation function $\Sigma(t)$, which is calculated perturbatively in powers of $A_e/\gamma_S B$.  The system decay time can therefore be extended by increasing the magnetic field, whereas the bath correlation time is fixed ($\tau_c=\tau_n\sim N/A_e$ -- see Eq. (\ref{eq:timescales})), allowing the Markovian limit to be reached for sufficiently large magnetic field (typically, for $\gamma_S B>A_e$).  In this limit, the spin decays exponentially
\begin{equation}
\left<S_+\right>_t = \left<S_+\right>_0 e^{i\gamma_S B t-t/T_2},
\end{equation}
with decay rate given explicitly by \cite{Coish2008}
\begin{align}
 \frac{1}{T_2} & = \frac{\pi}{6} f\left(\frac{d}{q}\right)\sum_j\left(\frac{\nu_j A_e^j I^j(I^j+1)}{\gamma_S B}\right)^2 \frac{A_e^j}{N},\label{eq:decoherence-rate}\\
f(r) & = \frac{1}{r}\left(\frac{1}{3}\right)^{2r-1}\frac{\Gamma(2r -1)}{[\Gamma(r)]^3},\quad
	r=\frac{d}{q} >1/2.\label{eq:geom-factor}
\end{align}
Here, $f$ is a geometrical factor, depending on the dot dimensionality $d$ and shape of the envelope function $\psi(r)\propto e^{-(r/r_0)^q}$, $\nu_j$ is the abundance of nuclear isotope $j$ with associated spin $I^j$ and hyperfine coupling constant $A_e^j$.  From Eq. (\ref{eq:decoherence-rate}), we find a typical decoherence time $T_2\gtrsim 1\,\mu\mathrm{s}$ for a two-dimensional GaAs quantum dot with Gaussian envelope function containing $N \simeq 10^{6}$ nuclear spins in the perturbative regime ($\gamma_S B\gtrsim A_e$).
\begin{figure}
 \includegraphics[width=0.99\columnwidth]{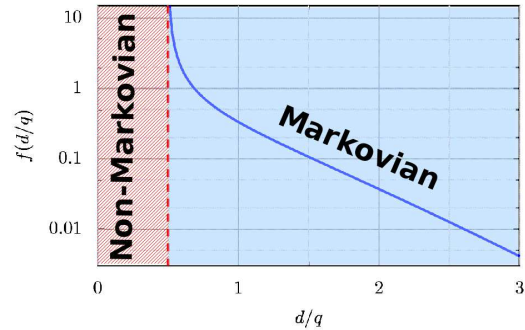}
\caption{\label{fig:geom-factor}
Geometrical factor from Eq. (\ref{eq:geom-factor}), determining the transverse-spin decay rate $1/T_2\propto f(d/q)$ for an isotropic electron envelope wave function of the form $\psi(r)\propto e^{\left(r/r_0\right)^q}$ in $d$ dimensions.  A Markov approximation is possible in the region labeled 'Markovian', where $f$ remains finite, but is not possible for $d/q \le 1/2$ ($d/q=1/2$ corresponding to a one-dimensional quantum dot with a Gaussian wave function).}
\end{figure}

The geometrical factor is plotted in Fig. \ref{fig:geom-factor}.  We note that the strong (exponential) dependence of $f$ on its argument for large argument suggests a strategy for reducing the decoherence rate $1/T_2$: increasing the dot dimensionality $d$ or decreasing the parameter $q$ with an appropriate confining potential.  Further, it will not be possible to perform a Markov approximation for any magnetic field $B$ when $d/q\le 1/2$, since the Markov approximation would result in a divergent rate $1/T_2\propto f(d/q)$.  The critical value  ($d/q = 1/2$) corresponds to a one-dimensional dot with parabolic confinement (resulting in a Gaussian envelope), which may be realized, e.g., in carbon nanotubes \cite{Churchill2008b}.

Here we have only discussed free-induction decay of electron spins, in the absence of refocusing pulses.  There has also been a substantial amount of work done on decay of the spin-echo envelope, both in theory \cite{DeSousa2003a,DeSousa2003b,Witzel2006,Yao2006,Yao2007,Liu2007,Shenvi2005a,Shenvi2005b,Cywinski2009} and experiment \cite{Koppens2008}.

\section{Nuclear-spin interactions and decoherence of heavy holes}\label{sec:hyperfine:hole}
%
%

In this section, we review recent studies on the interaction of a 
quantum-dot-confined heavy hole (HH) with the spins of the surrounding nuclei. 
Until very recently, HH were believed to interact only weakly with 
nuclear spins due to the $p$-symmetric HH
Bloch function \cite{Flissikowski2003, Shabaev2003, Woods2004, Bulaev2005a, Laurent2005,
Bulaev2007, Serebrennikov2007, Burkard2008}.
As we have seen in Sec. \ref{sec:hyperfine:electron}, a conduction-band electron 
couples to the nuclear
spins mainly via the Fermi contact interaction \eqref{eq:isotropichyperfine}. 
The coupling constants $A_k^e$ are proportional to the absolute-value-squared of the electron 
wavefunction at the site of the $k^{\mathrm{th}}$ nucleus. 
Only $s$-type orbital wavefunctions lead to a non-vanishing contribution
to the Fermi contact interaction because states of higher angular momentum ($p$, $d$, $\ldots$)
are represented by orbital wavefunctions which vanish at the positions of the nuclei.
On the other hand, the anisotropic hyperfine interaction (Eq. \eqref{eq:anisotropichyperfine}) 
and the coupling of electron orbital momentum to the nuclear spins 
(Eq. \eqref{eq:orbitalcoupling})
are irrelevant for conduction-band electrons because of the spherical symmetry of the orbital
wavefunction and the vanishing orbital angular momentum.
However, for valence-band electrons, and holes, the $p$-symmetry of the orbital wavefunction
leads not only to a vanishing of the Fermi contact interaction but also to a significant
enhancement of the anisotropic hyperfine interaction and the coupling of orbital
angular momentum. In the remainder of this section, we will discuss strongly
confined (quasi-two-dimensional) systems exhibiting a large splitting between HH
and light-hole (LH) states. This splitting is essential since it provides a well-defined
two-level system in the HH band.

\begin{figure}[t]
\centering
\includegraphics[width=.99\columnwidth]{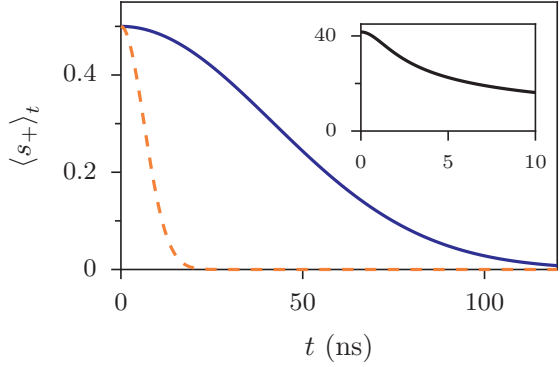}
\caption{\label{fig:hole-decoherence-outofplane} 
  Gaussian dephasing of a HH spin state transverse to a magnetic field
  $|\mathbf{B}|=B_z=10 \, \mathrm{mT}$ applied along the growth direction [001]
  of the crystal [solid line, Eq. \eqref{eq:hole-gaussian}]
  and dephasing of the transverse
  spin of a conduction-band electron under similar conditions (dashed line).
  Inset: HH coherence time $\tau_\perp$ (in ns) [Eq. \eqref{eq:hole-tau-perp}] as a function
  of the applied magnetic field (in T). The coherence time becomes shorter when increasing the
  external magnetic field because of the diamagnetic squeezing of the envelope wavefunction,
  effectively decreasing the number $N$ of nuclei in the quantum dot.
  An out-of-plane g-factor of $g_\perp \simeq 2.5$
  has been assumed \cite{vanKesteren1990}. Adapted from Ref. \cite{Fischer2008}.}
\end{figure}

It has been shown recently \cite{Fischer2008} that for HH the anisotropic hyperfine interaction 
and the coupling of orbital angular momentum to the nuclear spins can be rather strong, and that for 
unstrained quantum dots
\footnote{Strain-induced effects on the nuclear-spin interactions of confined holes have not yet been
considered theoretically. There is experimental evidence, however, that strong strain might lead to 
considerable mixing of HH and LH states \cite{Eble2008}, inducing non-Ising corrections to
Eq. \eqref{eq:hole-ising}.}, the effective interaction takes on a simple Ising form:
\begin{equation}
  \label{eq:hole-ising}
  H_{\mathrm{HH}} = \sum_k A^h_k \> s_z I_k^z.
\end{equation}
Here $A_k^h$ is the coupling of the HH to the $k^{\mathrm{th}}$ nucleus,
$s_z$ is the HH pseudospin-$\frac{1}{2}$ operator, and $I_k^z$ is the $z$-component
of the $k^{\mathrm{th}}$ nuclear-spin operator $\mathbf{I}_k$. 

The form of the interaction in Eq. \eqref{eq:hole-ising} is different from the Heisenberg 
interaction \eqref{eq:Zeeman-hf}
of electrons with the nuclear spins in a quantum dot and has profound consequences for the dynamics of quantum-dot-confined hole spins, as compared to electron spins.
If a magnetic field is applied perpendicular to the quantum dot, the transverse spin
dynamics (in the frame rotating with the frequency induced by $B_z$) are governed by 
a Gaussian decay (see Fig. \ref{fig:hole-decoherence-outofplane})
\begin{equation}
  \label{eq:hole-gaussian}
  \left\langle s_+ \right\rangle_t=\frac{1}{2}\exp{\left(-\frac{t^2}{2 \tau_\perp^2}\right)},
\end{equation} 
where $s_+ = s_x + i s_y$ and $s_j = \sigma_j/2$ with $\sigma_j$ being the $j^{\mathrm{th}}$ Pauli
matrix. The characteristic timescale is given by
\begin{equation}
  \label{eq:hole-tau-perp}
  \tau_\perp = \frac{1}{\sigma_h},
\end{equation}
where
\begin{equation}
  \sigma_h^2 \simeq \frac{1}{4 N} \sum_j\nu_jI^j(I^j+1)(A^j_h)^2
\end{equation}
is the variance of the random nuclear magnetic field `seen' by the HH. The sum runs over the
different nuclear isotopes in the dot and weights the coupling strengths $A^j_h$ 
of the HH with nuclear spins of species $j$ by the nuclear isotope abundances $\nu_j$. Furthermore,
$N$ is the number of nuclear spins in the dot and $I^j$ is the spin projection of the $j^{\mathrm{th}}$
isotope along the quantization axis (taken to be [001]).

\begin{figure}[t]
\centering
\includegraphics[width=.99\columnwidth]{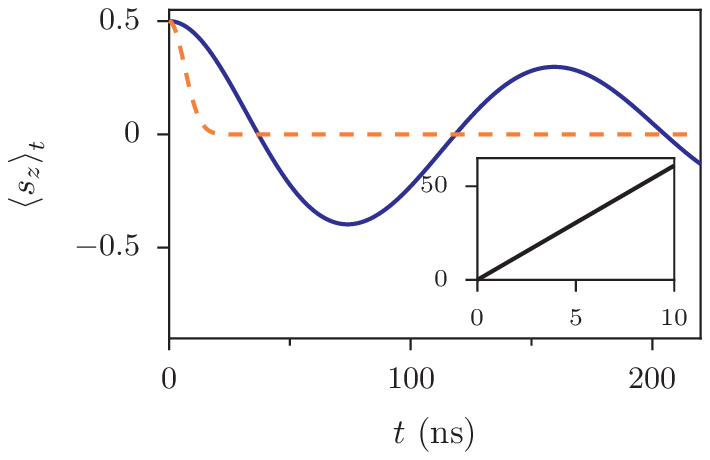}
\caption{\label{fig:hole-decoherence-inplane} 
  For a magnetic field applied in the plane of the quantum dot
  (here: $|\mathbf{B}|=B_x=10 \, \mathrm{mT}$), the transverse
  component of the HH spin decays according to a slow power-law [solid line, Eq.
  \eqref{eq:hole-powerlaw}], while the transverse spin of a conduction-band electron 
  under similar conditions shows a much faster Gaussian decay (dashed line).
  Inset: HH coherence time $\tau_\|$ (in $\mu \mathrm{s}$) [Eq. \eqref{eq:hole-tau-parallel}]
  as a function of the applied magnetic field (in T). 
  An in-plane g-factor of $g_\| \simeq 0.04$ has been assumed \cite{Marie1999}.
  Adapted from Ref. \cite{Fischer2008}.}
\end{figure}

If, however, a magnetic field is applied in the plane of the quantum dot, 
the transverse hole-spin component decays following a slow power-law decay
(see Fig. \ref{fig:hole-decoherence-inplane})
\begin{equation}
  \label{eq:hole-powerlaw}
  \left<s_z\right>_t\simeq\frac{\cos\left(b_\parallel t+\frac{1}{2}\arctan(t/\tau_\parallel)\right)}
  {2[1+(\frac{t}{\tau_\parallel})^2]^{1/4}}.
\end{equation}
The characteristic timescale of the decay is given by
\begin{equation}
  \label{eq:hole-tau-parallel}
  \tau_\parallel=\frac{b_\parallel}{\sigma_h^2},
\end{equation}
where $b_{\|}=g_{\|} \mu_B B_{\|}$, $g_{\|}$ is the in-plane HH g-factor, 
$\mu_B$ is the Bohr magneton, and $B_{\|}$ is the magnitude of the applied magnetic field.
We note that in two-dimensional GaAs systems, $g_\| \ll g_\perp$ for
HH \cite{Marie1999}.

Estimates for the coupling strengths $A_h^j$ in GaAs quantum dots have been given in
\cite{Fischer2008}: Weighting over the natural isotope abundances,
$\nu_{^{69}\mathrm{Ga}} \simeq 0.3$, $\nu_{^{71}\mathrm{Ga}} \simeq 0.2$, $\nu_{^{75}\mathrm{As}} 
\simeq 0.5$, the interaction strength has been found to be of order
\begin{equation}
  \label{eq:holecoupling}
  A_h = \sum_j \nu_j A_h^j \simeq -13 \, \mu e \mathrm{V}.
\end{equation}

This result indicates that the nuclear-spin interactions of a HH are only one order
of magnitude weaker than the nuclear-spin interactions of an electron, $A_e
\simeq 90 \mu e \mathrm{V}$ (see Sec. \ref{sec:hyperfine:electron}).
On the other hand, the functional form of the decay given in Eq. \eqref{eq:hole-powerlaw} provides
much longer coherence times as compared to the Gaussian decay of electron spins under similar
conditions (dashed line in Fig. \ref{fig:hole-decoherence-inplane}). 
Although there might be other mechanisms that limit the lifetime of hole-spin-state
superpositions (e.g., spin-orbit interactions), the coherence times associated with
nuclear-spin interactions can be as long as several tens of microseconds 
at typical laboratory magnetic fields (see inset of Fig. \ref{fig:hole-decoherence-inplane}).

An experimental verification of the predicted single-hole-spin coherence times has 
yet to be carried out.
However, recent experiments have shown very encouraging results which can be
regarded as prerequisites for single-hole-spin coherence-time measurements.
Ohno \textit{et al.} \cite{Ohno1999} have shown electrical hole-spin injection into 
a non-magnetic semiconductor and transport of the polarized holes over 200 nm.
Both Grbic \textit{et al.} \cite{Grbic2005}
and Komijani \textit{et al.} \cite{Komijani2008} were able
to fabricate gated quantum dots in a two-dimensional hole gas and to perform 
Coulomb-blockade measurements, indicating that the number of confined holes could 
be reduced at least close to the single-hole regime.
The capability to initialize and read out a single hole spin in self-assembled quantum 
dots by optical means has been reported by several groups 
\cite{Heiss2007, Ramsay2008, Gerardot2008, Eble2008}.
Ensemble hole-spin dephasing times of order nanoseconds have been measured 
in $p$-doped quantum wells by Syperek \textit{et al.} \cite{Syperek2007}.

Gerardot \textit{et al.} \cite{Gerardot2008} have measured the hole-spin injection fidelity
in self-assembled quantum dots. In this experiment, spin-polarized holes could be injected
into the dot with extremely high fidelity at low magnetic fields where nuclear-spin interactions
are expected to be dominant over spin-orbit interactions. This result indicates that nuclear-spin
interactions do not provide an efficient mechanism to flip the HH spin. In other words,
the spin states of the HH were only weakly coupled in the experiment, indicating an
Ising interaction as in Eq. \eqref{eq:hole-ising}.

Eble \textit{et al.} \cite{Eble2008} have recently reported on hole-spin relaxation in 
strongly strained, non-flat self-assembled InAs/GaAs quantum dots. 
Both the dot extension along the growth direction and the strong strain can lead to 
a considerable mixing of HH
and LH states, opening a new channel for hole-spin relaxation via the LH band.
It can be expected that there are significant corrections to the Ising Hamiltonian 
\eqref{eq:hole-ising}
due to this band hybridization. Accordingly, the reported relaxation times are quite short:
of order 15 nanoseconds. 


\section{Spin relaxation and decoherence in InAs nanowire-based quantum dots}
\label{sec:spinorbit}
%

\begin{figure}[t]
\centering
\includegraphics[width=.99\columnwidth]{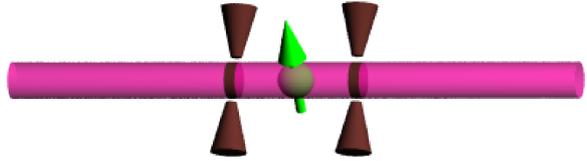}
\caption{Schematic picture of the system considered in Sec. \ref{sec:spinorbit}:
  a single-electron quantum dot is defined in an InAs nanowire via the potential created by the gates depicted in brown.}
\label{fig:InAs_dot}
\end{figure}

In this section we address the issue of relaxation and decoherence of an 
electron spin localized in an InAs nanowire quantum dot (QD),
schematically depicted in Fig. \ref{fig:InAs_dot}.   
Nanowire-based QDs are promising alternatives to the usual 
two-dimensional gate-defined QDs since there is only need of externally confining along 
one direction \cite{Bjork:04, Fasth:05, Pfund:06, Shorubalko:07}. 
One important difference between GaAs and InAs materials is the strength of spin-orbit 
interaction (SOI), this being much stronger in the latter. 
This has both positive and negative implications for using the electron spin as 
the fundamental unit of quantum computation (i.e., as a qubit): the strong SOI 
allows for control of the electron spin on very short timescales by means of electric 
fields as opposed to magnetic fields. At the same time, it implies  stronger coupling 
to the charge environment (e.g., phonons), which is expected to cause fast 
relaxation and decoherence of the electron spin. It is thus important to investigate 
the relaxation mechanisms in InAs quantum dots and to find means to suppress them, 
while allowing for a fast coherent control of the spin via electric fields. 

Relaxation and decoherence of the electron spin arise mainly from the coupling to 
the bath of phonons and the collection of nuclei in the QD. 
The phonon contribution has been studied microscopically 
in great detail for the case of gate-defined GaAs QDs in two-dimensional electron gases, 
and it has been shown that for large magnetic fields, similar to  the present case, the main 
contribution to relaxation comes from deformation-potential phonons,the associated 
decay time being of order $T_1\sim 10^{-4}-10^{-2}{\rm s}$ \cite{Vitaly2004}. As a 
consequence, a shorter  relaxation time can be expected for InAs QDs since 
the SOI is one order of magnitude larger than in GaAs ($T_1\propto (\lambda_{SO}/R)^2$, 
$R$ and $\lambda_{SO}$ being the radius of the wire and the spin-orbit length, 
respectively). However, as opposed to the bulk case, the phonon spectrum in 
nanowires becomes  highly non-trivial due to the mixing of the 
branches by the boundaries \cite{Norihiko1,Norihiko2}, leading to a strong 
modification of the relaxation time. We note that hyperfine-related effects are not
strongly affected by confinement, thus leading to similar dephasing times as in 
two-dimensional gate-defined QDs, see Ref.~\cite{trif2008}.

In cylindrical nanowires, there are three types of acoustic phonon modes: 
torsional, dilatational and flexural \cite{cleland}. 
These modes couple to the electrons in the nanowire and, in principle,
also to the electron spin via spin-orbit interaction. However, due to the special 
form of the spin-orbit interaction (i.e., linear in momentum), only a 
small fraction of the phonons in the spectrum mentioned above couples to the spin 
and gives rise to spin relaxation. 

In the following, we assume hard-wall boundary conditions for the electron 
confinement in the radial direction with the appropriate wave functions and 
energies \cite{trif2008}. We also assume that the extension of the wavefunction 
along the wire is much larger than the radius, i.e., $l\gg R$.  
The electron couples to the lattice (phonon field) via the deformation-potential 
interaction given by $H_{e-ph}=\Xi_0\nabla{\bm{u}}(\bm{r},t)$, 
where $\Xi_0$ is the deformation-potential strength and
\begin{equation}
\bm{u}(\bm{r},t)=\frac{1}{\sqrt{N}}\sum_{\bm{k}}[\bm{u}(\bm{k},\bm{r})b_{\bm{k}}(t)+{\rm 
h.c.}],\label{phononfield}
\end{equation}
with the displacement field $\bm{u}(\bm{k},\bm{r})$ given by \cite{Norihiko1,cleland}
\begin{equation}
\bm{u}(\bm{k},\bm{r})=\nabla\Phi_0+(\nabla\times\bm{e}_z)\Phi_1+(\nabla\times\nabla\times\bm{e}_z)
\Phi_2.\label{displacement}
\end{equation}
The index $\bm{k}\equiv\{q,n,s\}$ describes the relevant quantum numbers, i.e., 
the wave-vector along the wire, the winding number and the radial number, respectively, 
$b_{\bm{k}}(t)$ is the phonon annihilation operator, $\bm{e}_z$ is the unit 
vector along the $z$-direction (the symmetry axis of the wire) and
$\Phi_j=\chi_jf_{ns}^j(r)e^{i(n\phi+qz)}$, with $j=0,1,2$, $n=0,\pm1,\pm2,\dots$. 
The functions $f_{ns}^j(r)$ depend only on the radius \cite{Norihiko1,Norihiko2} 
and the $\chi_j$ are normalization factors. 

\begin{figure}[t]
\centering
\includegraphics[width=.99\columnwidth]{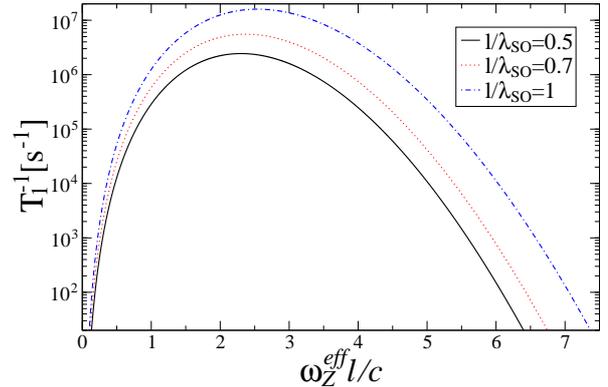}
\caption{The relaxation rate $T_1^{-1}$ as a function of the ratio $\omega_Z^{{\it eff}}l/c$ 
for three different ratios $l/\lambda_{SO}$ (see text). Adapted from Ref.~\cite{trif2008}}
\label{FigRelaxLong}
\end{figure}

Before investigating the spin relaxation in detail, we note that starting from the 
bare SOI Hamiltonian in \eqref{eq:spinorbit} and making use of the $\bm{k}\cdot\bm{p}$ 
method \cite{winkler} one can derive an effective SOI for an electron in the 
conduction band \cite{Fasth:07}:
\begin{equation}
H_{SO}=\gamma(\bm{k}\cdot\bm{c})(\bm{\eta}\cdot\bm{\sigma}).
\label{SO_ham}
\end{equation}   
Here, $\bm{\eta}=(\eta_x,\eta_y,\eta_z)$ is a vector of coupling constants \cite{Fasth:07}, 
$\bm{c}$ is the unit vector along the growth direction, $\bm{k}$ is the wave-vector of 
the electron along the wire, and $\bm{\sigma}$ is the electron-spin operator. 
In the presence of an external, static magnetic field applied along an arbitrary direction $\tilde{z} $, one can derive an effective 
spin-phonon coupling for the electron spin localized in the QD 
(for more details on the derivation see Ref. \cite{trif2008}):
\begin{equation}
H_{s-ph}=\frac{1}{2}g\mu_B\delta B_{\tilde{x}}(t)\sigma_{\tilde{x}}+\frac{1}{2}g\mu_B \delta 
B_{\tilde{z}}(t)\sigma_{\tilde{z}},\label{splong}
\end{equation}
with
\begin{equation}
\delta B_{\tilde{x},\tilde{z}}=B_{\it eff}\frac{\Xi_0}{\hbar\omega_0}\sum_{q,s}\frac{M_{s-ph}^{\tilde{x},\tilde{z}}(q)}{c_s^2\sqrt{2\mathcal{F}(q,s)\rho_c\omega_{q,s}/\hbar}}\omega_{q,s}^2 \bm{k} b_{\bm{k}}^{\dagger}+{\rm h.c.}, 
\label{fluctuatinglong}
\end{equation}
and $\bm{k}\equiv\{q,s\}$, $B_{\it eff}=B\exp{[-(l/\lambda_{SO})^2]}$, $\mathcal{F}(q,s)=\hbar R^2/4M\chi_0^2\omega_{q,s}$, where $\lambda_{SO}=\hbar/m|\eta|$ is the spin-orbit length, $M$ is the atomic mass, and $c_s$ is the speed of sound in the material. The functions $M_{s-ph}^{\tilde{x},\tilde{z}}$ are given explicitly in Ref.~\cite{trif2008}, but for small $q$ they scale as $M_{s-ph}^{\tilde{x}}\propto q$ and  $M_{s-ph}^{\tilde{z}}\propto q^2$. We mention that the phonons coupling to the spin in the lowest order  correspond to the branch $n=0$ in the phonon spectrum \cite{trif2008}.
We see that the SOI leads to both relaxation and pure dephasing of the electron spin. 
However, since the deformation-potential phonons are
superohmic (even in one dimension), the pure-dephasing rate vanishes \cite{Weiss}, so that 
in the following we can restrict ourselves to the first term in Eq. (\ref{splong}).
The relaxation rate $T_1^{-1}$ within the Bloch-Redfield approach  reads 
$T_{1}^{-1}={\rm Re}(J_{\tilde{x}\tilde{x}}(\omega_Z^{\it eff})+J_{\tilde{x}\tilde{x}}(-\omega_Z^{\it eff}))$,
with $\omega_Z^{\it eff}=g\mu_BB_{\it eff}/\hbar$ and
the correlation function $J_{ij}=(g\mu_BB/2\hbar)^2\int_0^{\infty}dt
\exp{(-i\omega t)}\langle\delta B_{i}(0)\delta B_j(t)\rangle$, leading to
\begin{equation}
T_{1}^{-1}\simeq \frac{\hbar}{4\pi\rho_cR^2l^3}\left(\frac{\Xi_0}{\hbar\omega_0}\right)^2\left(\frac{\omega_Z^{\it eff}l}{c_s}\right)^3[\mathcal{M}_{s-ph}^{\tilde{x}}(\omega_Z^{\it eff}l/c_s)]^2.\label{finalrelaxlong}
\end{equation}
In Fig. \ref{FigRelaxLong}, we plot the relaxation rate $T_{1}^{-1}$ as a function of the
dimensionless parameter $\omega_Z^{\it eff}l/c$ for different SOI strengths quantified 
through the ratio $l/\lambda_{SO}$. Here we have assumed $R=10\, {\rm nm}$ and  $l=50\, {\rm nm}$, which gives  
$\hbar c/l\equiv\hbar\omega_{ph}^l=0.05\,{\rm meV}$ and $\hbar c_l/R\equiv
\hbar\omega_{ph}^{R}=0.25\,{\rm meV}$.  We see in Fig. \ref{FigRelaxLong} that the relaxation rate 
$T_1^{-1}$ becomes rather large ($T_1^{-1}\sim 10^{5}-10^{7}\,{\rm s^{-1}}$) for $\omega_Z^{\it eff}/\omega_{ph}^l\sim 1-5$, i.e., when $\omega_Z^{\it eff}$ and  $\omega_{ph}^l$ are comparable. 

As mentioned before, one can make use of the strong SOI to coherently control 
the spin with electric fields. It was shown both theoretically \cite{vitaly:06} 
and experimentally \cite{nowack:07} that this is indeed possible on timescales of order 
nanoseconds.
The electric-dipole-induced spin resonance (EDSR) scheme proposed in 
Ref.~\cite{vitaly:06} suggests applying an alternating electric field $\mathcal{E}(t)$ to the 
QD, which, via electric dipole transitions and the SOI, gives rise to an effective 
alternating magnetic  field. If only the dipolar coupling to the alternating electric 
field $\mathcal{E}(t)$ is considered, we get $H_{e-el}(t)=e\mathcal{E}(t)y$, with the electric 
field $\mathcal{\bm{E}}(t)$ along the $y$-direction. This gives rise to an effective spin-electric-field coupling $H_{s-el}\equiv\delta B(t)\sigma_y$, with the time-dependent magnetic field 
\begin{equation}
\delta B(t)\sim e\mathcal{E}(t)R\frac{E_Z}{\hbar\omega_0}\frac{l}{\lambda_{SO}}e^{-l^2/\lambda_{SO}^2}.\label{EDSR}
\end{equation}
The electric field $\mathcal{E}(t)$ is assumed to have an  oscillatory behavior, 
$\mathcal{E}(t)=\mathcal{E}_0\cos{\omega_{ac}t}$, where $\omega_{ac}$ is the frequency of the ac 
electric field. By tuning the frequency of the oscillatory electric field $\omega_{ac}$ to 
resonance with the qubit Zeeman splitting $E_{Z}^{{\it eff}}$, one can achieve 
arbitrary rotations of the 
spin on the Bloch sphere on time scales given by the Rabi frequency $\omega_R=\delta 
B(0)/\hbar$ \cite{vitaly:06}. 
We mention that, within lowest order in the SOI, the induced fluctuating 
magnetic field $\bm{\delta{B}}(t)$ is always perpendicular to the applied field $\bm{B}$ and 
reaches its maximum when the applied electric field $\bm{\mathcal{E}}(t)$ points along 
the same direction as $\bm{B}$ \cite{vitaly:06}. 

We now give some estimates for the Rabi frequency $\omega_R$. We 
assume $l\approx 50 {\rm nm}$ and choose the amplitude of the electric field to be 
$\mathcal{E}_0\approx 10\, {\rm eV/cm}$. With these values we obtain 
$\omega_R\approx 10 \,{\rm GHz}$, which leads to timescales for the electron-spin control 
which are of order $\omega_R^{-1}\approx\, 0.1 {\rm ns}$. This timescale should 
be of course much shorter than the relaxation and decoherence times for the electron spin 
in the QD. Thus, in order to obtain a strong spin-electric-field coupling, there is need for a 
large effective Zeeman splitting $E_Z^{\it eff}\gg\omega_{ph}^l$. 
At the same time, one should keep the Zeeman splitting below the energy corresponding  to the next phonon
branch, since above it we find a substantial increase of the
relaxation rate. Since this next phonon branch ($n=1$) starts around
$2\hbar\omega_{ph}^R\approx 0.5\,{\rm meV}$, the condition for strong
spin-phonon coupling and weak relaxation becomes
$\hbar\omega_{ph}^l\ll E_Z^{\it eff}<2\hbar\omega_{ph}^R$. In this
regime, also the necessary condition $E_Z/\hbar\omega_0\ll1$ is satisfied, since 
for $l=50\,{\rm nm}$ we have $\hbar\omega_0=1.3 \, {\rm meV}$.  Note
the non-monotonic behavior of the relaxation rate as a function of the
effective Zeeman splitting (see Fig. \ref{FigRelaxLong}). This
non-monotonicity has the same origin as in GaAs QDs \cite{Vitaly2004}, and it
comes from the fact that for increasing Zeeman splitting the
wave-length of the phonon decreases. When this wavelength becomes less than
the length of the quantum dot, the phonons decouple from the electron (i.e., the
electron-phonon coupling averages to zero). A similar non-monotonic
effect has been recently observed in GaAs double QDs \cite{meunier:07}.

\section{Summary}\label{sec:summary}
%
%

We have discussed spin decoherence of electrons and holes confined to 
quasi-two-dimensional quantum dots and interacting with the surrounding 
nuclear spins, as well as spin relaxation of electrons in one-dimensional
nanowire quantum dots due to spin-orbit interaction.

Electrons can exhibit a well-defined exponential decay of their spin coherence,
with coherence times of order microseconds -- the timescale where virtual
flip-flops between electron and nuclear spins become important. 
This prediction relies on the assumption that the nuclei can be prepared in a
so-called `narrowed' state, which results in a suppression of field fluctuations
induced by the nuclear spins. 
Recent experiments indicate that this kind of nuclear-state preparation might soon be 
achievable also experimentally.

Heavy holes show a different decoherence behavior than electrons. For quasi-two-dimensional
unstrained quantum dots, the interaction with nuclear spins has 
a simple Ising form, in contrast to the Heisenberg interaction of electrons. 
By applying moderate magnetic fields in the plane of the quantum dot, coherence times
much longer than for electrons could potentially be achieved. 
The Ising interaction also implies that no hole-nuclear flip-flop processes are
present. This would make state narrowing even more efficient than for electrons, since
its beneficial effect would not be limited by the timescale set by flip-flop processes,
but rather by the (potentially much longer) timescales set by other mechanisms like 
nuclear dipole interactions. The Ising interaction also implies that spin relaxation due to
nuclear-spin interactions is a highly inefficient process.

As we have seen, spin-orbit-induced effects can be expected to be the main source of 
spin relaxation for electrons in InAs nanowire quantum dots.
The associated relaxation rate has an interesting non-monotonic functional
dependence on the effective Zeeman splitting and varies over
several orders of magnitude.

Quantum-dot-confined electrons and holes are excellent candidates for quantum-information
storage and processing devices. From a theoretical point of view, the results presented
in this article indicate that the use of holes rather than electrons might be advantageous because
of considerably longer spin coherence times. However, manufacturing gated single-spin qubits
in two-dimensional hole gases, as well as initializing, coherently manipulating and
measuring single hole spins, are still experimentally very challenging tasks.

%
%

We acknowledge funding from the Swiss NSF, NCCR Nanoscience, JST ICORP, QuantumWorks,
and an Ontario PDF (WAC).


\end{document}